# Field electron emission characteristic of graphene

Weiliang Wang, Xizhou Qin, Ningsheng Xu, and Zhibing Li[*]

State Key Lab of Optoelectronic Materials and Technologies,

and School of Physics and Engineering, Sun Yat-sen University,

Guangzhou 510275, People's Republic of China

**Abstract**

The field electron emission current from graphene is calculated analytically on a semiclassical model. The unique electronic energy band structure of graphene and the field penetration in the edge from which the electrons emit have been taken into account. The relation between the effective vacuum barrier height and the applied field is obtained. The calculated slope of the Fowler-Nordheim plot of the current-field characteristic is in consistent with existing experiments.

## 1. Introduction

The cold field electron emission (CFE) as a practical microelectronic vacuum electron source, that is driven by electric fields of about ten volts per micrometer or less, has been demonstrated by the Spindt-type cathodes, which is basically micro-fabricated molybdenum tips in gated configuration [1] . In recent years, much interest has turned to the nano-structures, such as the carbon nanotubes and nanowires of various materials [2, 3], for that the high aspect ratios of these materials naturally lead to high field enhancement at the tips of the emitters thereby lower the threshold of macroscopic fields for significant emission.

So far, most of the experimental efforts and theoretical studies on the possible applications and the physical mechanism of CFE have been concentrated on the quasi one-dimensional structures, such as carbon nanotubes and various nanowires. However, the CFE from two-dimensional structures should be worthy of a deep investigation as its current-field characteristic is completely different from the conventional Fowler-Nordheim (FN) law [4]. Since the experimental realization of the free-standing graphene, [5] that is a mono-layer of carbon atoms packed into a hexagonal two-dimensional lattice, the two-dimensional atomic crystal has aroused great interest both in experimental and theoretical studies. Its one-dimensional edge of atomic thickness is a unique feature and of particular interest. Graphene has an excellent electrical conductivity, as an attractive CFE emitter should be. Several groups have demonstrated that graphene does show promising CFE properties, such as a low emission threshold field and large emission current density [6-12]. The present article should study the CFE of graphene theoretically.

---

[*] Email: stslzb@mail.sysu.edu.cn

The conventional FN theory for the characteristic relation between the CFE current and the applied macroscopic electric field was derived for the planar emitters in principle, though it has been known that the FN theory is also qualitatively fitted to the results of most CFE experiments of nano-structures. On the other hand, sophistic theoretical studies of CFE of nano-structures based on quantum chemistry methods [13-24] and electronic band structures [15] do reveal new mechanism and characteristics that are proprietary in nano-strutures. A classical study has indicated that the field emission of nano-sheet follows the $ln(J/F^3) \sim 1/F^2$ law in contrast to the conventional FN law of $ln(J/F^2) \sim 1/F$. S. Watcharotone et al. numerically obtained the field enhancement factor on the corners and edges of graphite sheets by employing the boundary element method [25]. It is still a challenge to find out the CFE characteristic that is rooted in the unique microscopic electronic structure of graphene. Instead of treating graphene as an ideal metallic sheet, the present paper should take into account the energy band structure and density of states of graphene. More importantly, the field penetration should be considered, as the field penetration would play decisive role in the CFE of nanostructures [24, 26]. Based on a self-consistent semiclassical model, we will give the charge distribution (and thus the field penetration) on graphene under the macroscopic applied field analytically. As argued in the conclusion that the emission current from the armchair edge is much stronger than that from the zigzag edge, therefore we should only consider the CFE of the armchair edge.

The model will be described in the section 2, where contains the well-known band structure of graphene and a phenomenological semi-classical model for the charge accumulation under the applied field. The section 3 presents the solution of the charge distribution and the effective vacuum barrier under the applied field. The characteristic emission current linear density versus applied field is given numerically in the section 4. The last section are the conclusions and discussions.

**2. Model**

We consider a planar graphene mounted on a metallic substrate (the cathode) vertically. The plane of graphene is parallel to the external applied electric field. For simplicity, the graphene is assumed to have a long upper edge such that the effects of its lateral edges could be ignored. The set-up of the graphene emitter is illustrated in Figure 1. The box is unphysical but used to guide the eyes. The height of the graphene is denoted by $H$. The macroscopic electric field $F$ is applied parallel to the graphene plane that is in the vertical direction of the figure. The local field along the upper edge of the graphene is enhanced as the electrons accumulate along the edge. Therefore the vacuum potential in front of the edge is greatly reduced and the electrons could have considerable probability to emit into the vacuum in the forward direction by quantum tunneling. The forwards emission is assumed in the following calculation.

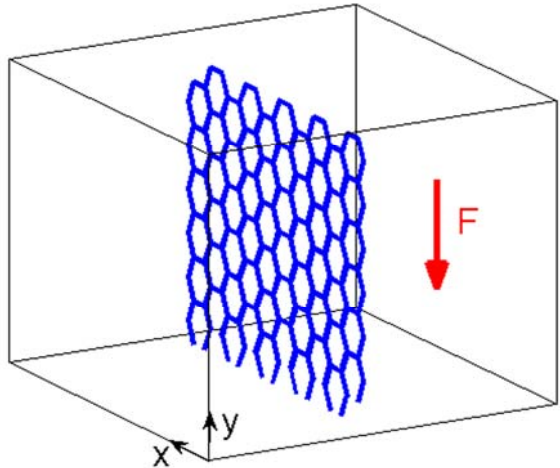

Fig. 1 (Color online) Structure of graphene with armchair edge as the upper edge.

The electronic properties of graphene can be found in many publications, for instance Ref. [27]. Here we collect the properties that are relevant to our topic. The unit cell of graphene as well as the first Brillouin zone are plotted in Fig.2a/b. The edges parallel to the y-axis are called Z-edges, while the edges paralleled to the x-axis are armchair edges.

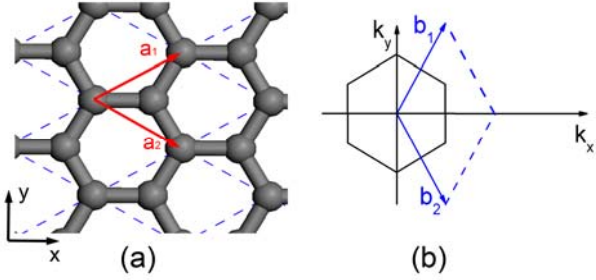

Fig. 2 (Color online) (a) Lattice structure of graphene, and $\vec{a}_1$ are $\vec{a}_2$ the lattice unit vectors. (b) Cooresponding first Brillouin zone, $\vec{b}_1$ and $\vec{b}_2$ are the reciprocal unit vectors.

According to the tight binding model, the dispersion relation of graphene is [28]

$$E_k = E_{f0} \pm t\sqrt{4\cos\frac{\sqrt{3}k_x a}{2}\cos\frac{k_y a}{2} + 2\cos k_y a + 3} \qquad (1)$$

Where $t$ ($\approx$2.8 eV) is the nearest-neighbor hopping energy, $a$=0.246nm is the magnitude of **a₁** and **a₂**, $E_{f0}$ is the intrinsic Fermi level (i.e., the energy level of the neutral graphene). The plus/minus sign corresponds to the conduction/valence band. The band structure is plotted in Fig.3, where the vertical axis is $(E_k-E_{f0})/t$. The enlarged part ($|E_k-E_{f0}|$<1.eV) is the K point region where the dispersion relation is linear and the electron can be described by the Dirac equation.

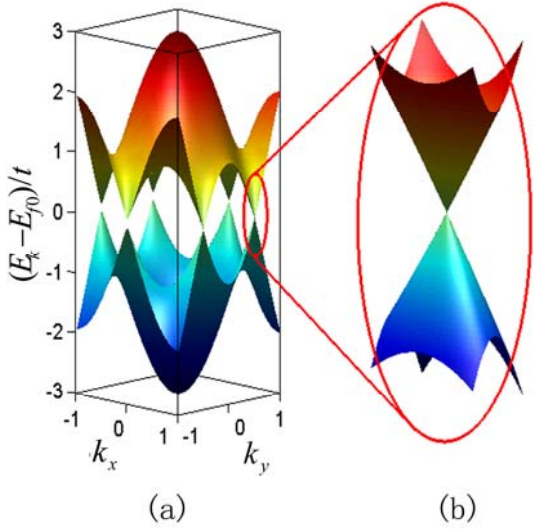

Fig. 3 (Color online) (a) Electronic energy band structure of graphene. (b) Zoom in of the dispersion relation close to the K point. The vertical axis is $(E_k - E_{f0})/t$. The wave vector scale is $2\pi/\sqrt{3}a$ ($4\pi/3a$) for $k_x$ ($k_y$).

In the K-point region, the density of states (DOS) in unit area of graphene is

$$D = \frac{2}{\pi(\hbar v_F)^2}\left|E - E_{f0}\right| \qquad (2)$$

Where the Fermi velocity is $v_F = \sqrt{3}at/2\hbar$. There are edge states at the intrinsic Fermi level along the Z-edge [29]. The density of edge states per length of edge is $n_e = 1/3$. The edge is neutral in average when the edge states are half-filling. In the last section, we will argue that the field emission from the Z-edge is much difficult than that from the armchair edge (A-edge). Therefore, from hereafter the emission edge is specified to be the A-edge. The A-edge emission is dominated by the states near the K-point for which the effective barrier height is just the work function.

The applied macroscopic field will induce charge in the graphene which in turn will alter the electrostatic potential. The charge density and the electrostatic potential should be determined self-consistently via the Possion equation and the Fermi-Dirac distribution in the energy levels given by the tight-binding theory (Eq.(1)). Each single/doubled occupied level which is higher than the intrinsic Fermi level ($E_{f0}$) contributes one/two negative unit charge/s (-e). Removing one/two electron/s from a level which is lower than $E_{f0}$ contributes one/two positive unit charge/s (e).. Under the quasi-equilibrium hypothesis, the real Fermi level ($E_f$) is fixed by the substrate that is set to be zero. The intrinsic Fermi level is varying with the electron potential energy ($u$) as $E_{f0}=E_f+u$ in space as we have ignored the variation of the exchange and correlation energies. At the zero temperature, the energy levels between $E_f$ and $E_{f0}$ are filled if $E_{f0}<E_f$. Else if $E_{f0}>E_f$, the energy levels between $E_f$ and $E_{f0}$ are empty.

The charge distribution is modulated via the electric energy potential. With the tight-binding DOS (2), one has spatial varying charge area density

$$\rho(z) = \frac{e}{\pi(\hbar v_F)^2} u(z) |u(z)| \qquad (3)$$

Since the external applied field exerts an upward force on the electrons, most induced charges are accumulated in the region near the upper edge, where the electrons are stopped by the vacuum energy barrier. As a consequence of the charge accumulation, the field in the graphene is more or less screened and the local field in the vacuum in the vicinity of the upper edge is enhanced. As an approximation, we assume that all induced charges (space charges) are distributed continuously in a strip of width $w$ by the end of the graphene, uniform in the direction paralleled to the A-edge. The region of space charge is illustrated in red in Fig.4. It is assumed that the length of the strip is so long that the lateral edge effects can be ignored.

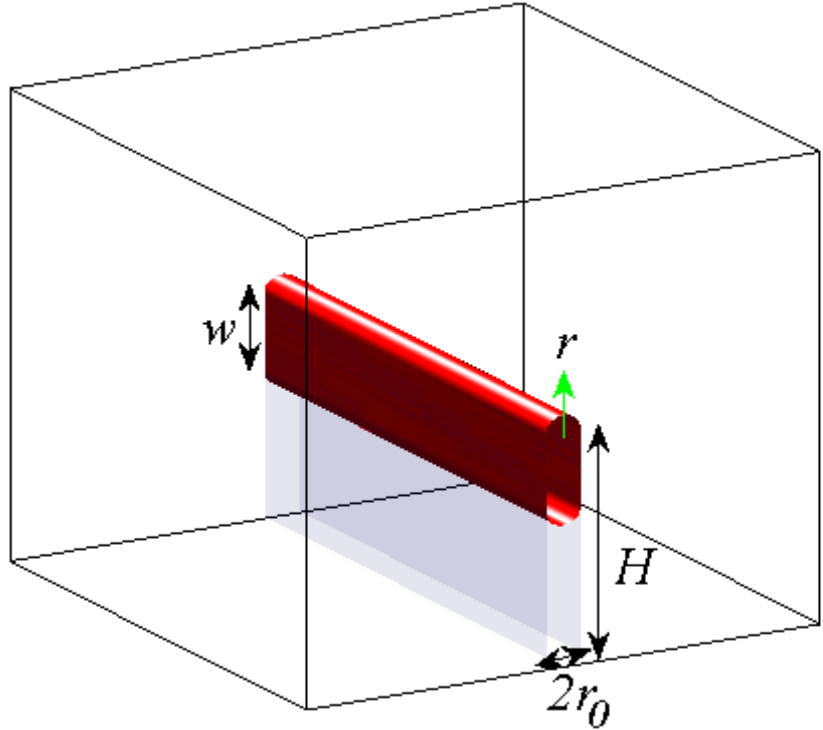

Fig. 4(Color online) Schematic illustration of the graphene emitter (gray). The bottom plane is the substrate. The red strip is the space charge region, with width $w$ and thickness $2r_0$.

We further introduce a quadratic ansatz for charge area density,

$$\rho(z) = -eA(z - H + w)^2 \tag{4}$$

Where $A$ and $w$ are two parameters to be determined. The boundary condition $\rho(H - w) = 0$ has been considered. The charges in the strip create a potential variation across the substrate plane. To remove this, we insert an image strip with charge area density $-\rho$ at distance $H$ below the substrate plane.

**3. Charge distribution in the graphene and the energy potential**

With the ansatz (4), the potential energy in the vacuum in vicinity of the upper edge of the graphene, reads

$$u(z) = -eFz - \frac{e^2 A}{4\pi\varepsilon_0} \int_{H-w}^{H} dz'(z'-H+w)^2 \log\frac{(z-z')^2 + r_0^2}{(z+z')^2 + r_0^2} \tag{5}$$

At $z = H$, (5) gives

$$u(H) = -eFH + \frac{e^2 w^3}{36\pi\varepsilon_0}\left(11 + 6\log\frac{2H}{w}\right) A \tag{6}$$

As we are interested in the case that $|u(H)| \ll eFH$, the l.h.s. can be ignored, so

$$A = \frac{36\pi\varepsilon_0 FH}{ew^3\left(11 + 6\log\frac{2H}{w}\right)} \tag{7}$$

The forwards derivative of the potential energy at the edge is

$$u'(H) = -eF - \frac{e^2 w^2 A}{4\pi\varepsilon_0}\left(2\log\left(\frac{w}{r_0}\right) - 3\right) \tag{8}$$

To determine $w$, we equate (8) to the classical result of an ideal metallic sheet of thickness $r_0$, $-e\gamma F$, where $\gamma = \frac{1}{2}\sqrt{\frac{\pi H}{r_0}}$ is obtained by the conformal mapping method. [4] When $H \gg w$ (it is self-consistent with its deduction), the width of the space charge strip (i.e., the penetration depth) can be estimated as (see Appendix A)

$$w = 6\frac{[1 + 3\log(2H/r_0)]B - 1}{[1 + 3\log(2H/r_0)]B + 1}\sqrt{\frac{r_0 H}{\pi}} \tag{9}$$

Where $B = 0.138$ is obtained by the numerical solution at $2H/r_0 = 10^5$. Note that the penetration depth given by $w$ is independent of the applied field in the high field regime of present discussion.

Because $u(H)$ is a lower order term compared to $eFH$, one can not obtain it by inserting (7) and (9) into (6). Instead, one should first evaluate $\rho^0(H) = -eAw^2$, then use the relation (3) to derive $u(H)$. The difference between the intrinsic Fermi level and the real Fermi level at the edge $E_{f0} - E_f = u(H) = \hbar v_F w\sqrt{\pi A}$. Using (7),

$$u(H) = -6\pi\hbar v_F \sqrt{\frac{\varepsilon_0 FH}{ew(11 + 6\log(2H/w))}} \tag{10}$$

Since $w$ is independent of F, $u(H) \sim -\sqrt{F}$. The Fermi level $E_f$ is fixed in the quasi-equilibrium,

the intrinsic Fermi level moves downwards as the applied field increases and thus the accumulated charge density increases. Choosing $r_0$=0.08 *nm* [20], our estimation gives $E_f$- $E_{f0}$ ~ 0.6 eV when F=16. V/μm and H=10. μm. Therefore the hypothesis of linear dispersion relation for graphene used in our previous model is consistent.

Once the charge distribution is known, the potential energy in the entire space, $u(\vec{r})$, can be easily calculated. The transmission coefficient and the emission current are studied in the next section.

**4. Emission current**

Generally, the emission current is given by

$$J = e\int_{-\infty}^{+\infty} T(E_\perp) n(E_\perp) dE_\perp \qquad (11)$$

where $E_\perp = E_k - W_{//}$ is the electron's normal energy (some times called its "forwards energy"), that is the electron total energy ($E_k$) subtracted by the kinetic energy of motion in the direction paralleled to the the edge ($W_{//}$); $T(E_\perp)$ is the tunneling probability of the k-state; $n(E_\perp)$ is the supply function, i.e., the number of electrons hits the potential barrier in unit time. The supply functions for the graphenes with the A-edge will be given in Appendix B.

The tunneling probability in the JWKB approximation reads

$$T(E_\perp) = \exp\left\{-2\int_{r_a}^{r_b} \sqrt{\frac{2m}{\hbar^2}\left[W_{eff} + u(r) - e\varphi_{im}(r)\right]}\, dr\right\} \qquad (12)$$

where $r_a$/$r_b$ are the inner/ outer boundary of the unclassical region of the potential barrier. The effective barrier height $W_{eff} = W_0 - E_\perp + E_{f0}$ with $W_0$ is the work function of graphene in absent of applied field. The electrostatic potential $u(r)$ has been discussed in last section. The image potential $\varphi_{im}(r)$ is cut off to zero at $r=r_0$. Our previous *ab initio* simulation suggested that the image potential of the emitting electron, $\varphi_{im}(r)$, can be modeled as the image potential of a metallic sphere of radius $r_0$, [20]

$$\varphi_{im}(r) = \frac{er_0}{8\pi\varepsilon_0 (r^2 - r_0^2)} \qquad (13)$$

The virtual metallic sphere has radius of atom size.

As an example, the emission current linear density (current per length of edge) is calculated numerically for the graphene with the A-edge, with the graphene height $H=10.\mu m$ and the cut off parameter $r_0 = 0.08$ *nm*. The work function for the A-edge terminated by H is between 4.19 eV and 4.41 eV, according to our ab initio simulation to be given elsewhere. The characteristic *J-F* diagram and the Fowler-Nordheim (FN) plot are presented in Fig.5 for the lower and upper limit of work function.

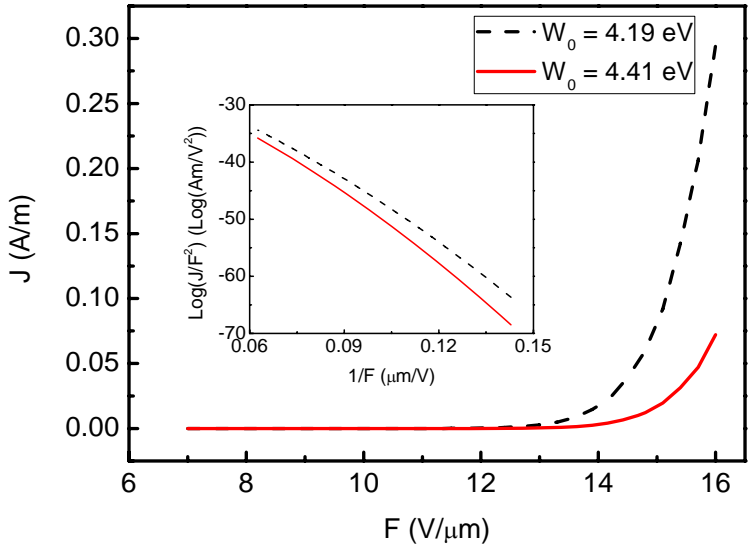

Fig. 5 (color online) J-F characteristic of the graphene with the armchair edge. The inset is the corresponding FN plot. ($H$=10. μm, $r_0$ = 0.08 nm)

The FN plot is not a strict straight line. The slope of the FN plot is about $-3\times10^8 m/V$ for the typical emission current $J \approx 0.1 A/m$, which is close to the results of experiments[6, 8, 10, 11, 30]. Table 1 lists the slopes of the FN plots of the experiments. Some experimental results are significantly smaller (absolute value) than ours. We attribute this discrepancy mainly to the edge roughness in the corresponding experiments, the uncertainty of the graphene height, and maybe the corner effect.

Table 1. Slope of FN plots (Ref. A [11], Ref. B [8], Ref. C [30], Ref. D [6], Ref. E [10])

|         | This model        | Ref. A            | Ref. B            | Ref. C              | Ref. D            | Ref. E            |
|---------|-------------------|-------------------|-------------------|---------------------|-------------------|-------------------|
| Slope (V/m) | $-3 \times 10^8$ | $-1 \times 10^8$  | $-6 \times 10^7$  | $-1.3 \times 10^7$  | $-1.6 \times 10^8$ | $-4 \times 10^8$  |
|         |                   | $-2.3 \times 10^7$ | $-2.5 \times 10^7$ |                     | $-6 \times 10^7$  | $-1 \times 10^8$  |

## 5. Conclusion and discussions

We proposed the phenomenological semi-classical model for the field electron emission from graphene that incorporates the energy band structure of graphene. The charge distribution and the electrostatic potential are related by the classical Poisson equation under the restriction of the quantum mechanical density of states. When the external field is applied paralleled to the graphene plane, electrons accumulate in the end edge of the graphene. The local field along the edge is thereby enhanced due to the screening effect. At the same time, the effective vacuum barrier height relative to the Fermi level is lowered, or equivalently the real Fermi level rises with respective to the top of the vacuum barrier.

The present paper only considered the CFE from armchair edges. In the following, we argue that the CFE from Zigzag edges could be ignored. According to the energy band structure of graphene, it is mainly the states near the K-point, whose energy dispersion is linear and has no gap, and the edge states, that are occupied when the field is applied to the graphene. The tunneling probability depends only on the forward emission energy ($E_\perp$) which is given by $E-E_{//}$, with $E \sim E_f$ the total energy and $E_{//}$ the parallel kinetic energy. The absolute vacuum energy barrier height is above the intrinsic Fermi level by the work function $W_0$ in the absent of applied field (here the image potential is ignored). The effective barrier height for the band states near the realistic Fermi level is ([4, 31] for general discussions)

$$W_{eff} = W_0 + E_{f0} - E_f + E_{//} \tag{18}$$

For the emission from the aimchair, the minimum of $E_{//} \sim 0$, therefore the effective barrier height is $W_0 + E_{f0} - E_f$ that is smaller than $W_0$ and decreasing with the applied field (see (10)). On the other hand, for the emission from the Z-edge, the minimum of $E_{//} \sim t$. Therefore, the zero-field barrier height is higher than the work function by *t*, i.e., is about 7.0eV. Another fact that disbennifits the CFE from the Z-edge is that the Z-edge has edge states with energy near the intrinsic Fermi level. That will pin the Fermi level at $E_{f0}$, that is, $E_f - E_{f0} \sim 0$. So the effective barrier height does not decrease significantly in the applied field. Thus the emission current density from the zigzag edge is much smaller than that from the armchair edge.

Now come back to the armchair CFE. The varying of the potential energy barrier height versus the applied field, which is the key parameter related to the field penetration, is given analytically in the present paper. It is found that the barrier height seen by an electron with Fermi energy can be significantly lowered by the applied macroscopic field. The difference between the work function and the barrier height is proportional to the square root of the applied field. In the experimental range of field, the barrier height can decrease more than half eV. That is an important feature responsible to the effective field emission of graphene. We also find that screening depth of the graphene is almost independent of the applied fields. It is proportional to the square root of the height of the graphene.

Taking the band structure of graphene into account, we calculated the supply function of graphene with the armchair edge. Thus the *J-F* characteristic curve and FN plot are obtained. The slope of the FN plot of field emission from the armchair edge is close to those measured in the recent experiments.

In the following, we discuss the possible errors and uncertainty of the model.

To determine the parameters of the ansatz of the charge density, the residue potential drop at the edge is assumed to be much smaller than the unscreened voltage drop of the macroscopic applied field. That is the case when the applied field is high. When the applied field is week, the higher order correction should be considered.

Our model is for an ideal flat mono-layer graphene with well-defined armchair edge. In most experiments, instead, a bundle of graphene is used as emitter. There is screening effect between graphene sheets. The graphene sheets are not exactly perpendicular to the substrate. These uncertainties would be accounted by an effective applied field and an effective graphene height thereby our model is still valid qualitatively. The roughness of edge and the finite length of edges have not been incoperated in the present paper. They should be important and interesting, as addressed in Refs. [25, 32].

**ACKNOWLEDGMENTS**

The project is supported by the National Natural Science Foundation of China (Grant Nos. 10674182, 10874249, and 90306016), the National Basic Research Program of China (2007CB935500 and 2008AA03A314) and the China Postdoctoral Science Foundation (20100470974).

**Appendix A: The penetration depth**

Equate (8) to $-e\gamma F$, and substitute $\gamma$ by $\frac{1}{2}\sqrt{\frac{\pi H}{r_0}}$, which is the field enhancement factor of ideal metallic sheet of thickness $r_0$, one has

$$-eF - \frac{e^2 w^2 A}{4\pi\varepsilon_0}\left(2\log\left(\frac{w}{r_0}\right) - 3\right) = -e\gamma F \qquad (A1)$$

Insert (7) and into (A1),

$$1 + \frac{9H}{w}\left(11 + 6\log\frac{2H}{w}\right)^{-1}\left(2\log\left(\frac{w}{r_0}\right) - 3\right) = \frac{1}{2}\sqrt{\frac{\pi H}{r_0}} \qquad (A2)$$

Define $\beta = 2H/r_0$ and $\alpha = w/r_0$, (A2) can be written as

$$\log\alpha = \frac{11}{6} + \log\beta + \frac{1 + 3\log\beta}{2\alpha/\beta - 3 - \alpha\sqrt{\pi/2\beta}} \qquad (A3)$$

We only consider the case that $2\alpha/\beta = w/H$ is much smaller than 1, such that it can be ignored. Then,

$$\log\alpha = \frac{11}{6} + \log\beta - \frac{2 + 6\log\beta}{6 + \alpha\sqrt{2\pi/\beta}} \qquad (A4)$$

Let $\alpha = s\beta^{1/2}/\sqrt{2\pi}$, where s is presumably a slow function of β, such that $\frac{ds}{sd\beta} \equiv \frac{\dot{s}}{s} \sim 0$. From (A4),

$$\frac{\dot{s}}{s} = \frac{1}{2\beta} - \frac{6}{\beta}\frac{1}{6+s} + \frac{2+6\log\beta}{(6+s)^2}\dot{s} \tag{A5}$$

Let it equal to zero, one has

$$\frac{1}{(6-s)(6+s)}ds = \frac{1}{4\beta(1+3\log\beta)}d\beta \tag{A6}$$

Integral of (A6) gives

$$s = 6\frac{(1+3\log\beta)B - 1}{(1+3\log\beta)B + 1} \tag{A7}$$

Where $B$ is an integral constant. Therefore,

$$\frac{w}{r_0} = \frac{6\sqrt{\beta}}{\sqrt{2\pi}}\frac{(1+3\log\beta)B - 1}{(1+3\log\beta)B + 1} \tag{A8}$$

Substitute $\beta = 2H/r_0$ in it,

$$w = 6\frac{[1+3\log(2H/r_0)]B - 1}{[1+3\log(2H/r_0)]B + 1}\sqrt{\frac{r_0 H}{\pi}} \tag{A9}$$

We fix $B$ by solving (A4) numerically for a specific β. A reasonable parameter for $\beta=2H/r_0$ is $10^5$. The corresponding $\alpha$ is $\alpha$=500.6, that leads to $B = 0.138$.

### Appendix B : Supply function for the armchair edge

With the energy dispersion relation (1), the supply function can be written as

$$n(E_y)dE_y = \int_{E_y} \frac{2}{1+\exp[\beta(E_k - E_f)]}\frac{1}{(2\pi)^2}\frac{\partial E_k}{\hbar\partial k_y}dk_y dk_z$$
$$= dE_y \int_{E_y} \frac{2}{1+\exp[\beta(E_k - E_{f0} + u(H))]}\frac{1}{(2\pi)^2 \hbar}dk_z \tag{B1}$$

where $\frac{\partial E_k}{\hbar\partial k_y} = v_y$ is the electron's velocity normal to the armchair edge, $\beta = 1/k_B T$, $k_B$ the Boltzmann factor, and $T$ the temperature. The potential energy at the edge, $u(H)$, is given by (10). As the normal direction is Y axis (Fig. 1 (a)) for the A-edge, the electron's normal energy is

$$E_\perp = E_y = E_k - W_{//}, \text{ with } W_{//} = \frac{\hbar^2 k_x^2}{2m}.$$

At the zero temperature, the Fermi-Dirac distribution is a step function. One simply has

$$n(E_y)dE_y = \begin{cases} \dfrac{dE_y}{2\pi^2\hbar} \displaystyle\int_{E_y} dk_x & E_k - E_f < 0 \\ 0 & E_k - E_f > 0 \end{cases} \quad (B2)$$

The integral is subjected to two constraints. The first is the energy constraint that requires $E_k - E_f < 0$. The second comes from the band structure that the states do not exist in certain range of $k_x$ when $E_y$ is given.

(1) Energy constraint

Denote $\varepsilon_y = E_y - E_{f0}$, $\Delta_y = \sqrt{\dfrac{2m}{\hbar^2}|\varepsilon_y|}$, and $\Delta_H = \sqrt{\dfrac{2m}{\hbar^2}|\varepsilon_y + u(H)|}$. For the states in the conduction band and lower than the Fermi level, $E_{f0} \leq E_k \leq E_f$. It turns out to be $0 \leq |k_x| \leq \Delta_H$ for $0 \leq \varepsilon_y \leq -u(H)$ and $\Delta_y \leq |k_x| \leq \Delta_H$ for $\varepsilon_y < 0$. For the valence band electrons, the energy constraint gives $0 \leq |k_x| \leq \Delta_y$.

(2) Band constraint

Introduce $q = k_y - \dfrac{4\pi}{3a}$. Near the K-point, one can expand the dispersion relation (1) for small $|k_x|$ and $q$, leading to

$$\varepsilon_y + W_{//} = \pm \hbar v_F \sqrt{k_x^2 + q^2} \quad (B3)$$

It can be written as

$$q^2 = \dfrac{1}{(\hbar v_F)^2}(\varepsilon_y + W_{//})^2 - k_x^2 \quad (B4)$$

It is clear that $q$ does not always have real solution for given $\varepsilon_y$ and $k_x$. With (B4) and a practical condition $|\varepsilon_y| \ll mv_F^2 \sim 5.6$ eV, the constraint that guarantees a real $q$ can be written as

$$|k_x| \leq \dfrac{|\varepsilon_y|}{\hbar v_F} \quad (B5)$$

Two constraints are combined to give

$$|k_x| \leq \min\left(\Delta_H, \frac{|\varepsilon_y|}{\hbar \upsilon_F}\right) \quad (B6)$$

for electrons in the conduction band and $\varepsilon_y \geq 0$; and

$$|k_x| \leq \min\left(\Delta_y, \frac{|\varepsilon_y|}{\hbar \upsilon_F}\right) = \frac{|\varepsilon_y|}{\hbar \upsilon_F} \quad (B7)$$

for electrons in the valence band and $\varepsilon_y < 0$.

The integral of (B2) gives

$$n(E_y) = \begin{cases} \dfrac{2}{\pi^2 \hbar} \dfrac{E_{f0} - E_y}{\hbar \upsilon_F}, & E_y < E_{f0} \\ \dfrac{2}{\pi^2 \hbar} \min\left(\Delta_H, \dfrac{E_y - E_{f0}}{\hbar \upsilon_F}\right) & E_{f0} \leq E_y < E_{f0} - u(H) \end{cases} \quad (B8)$$

The first (second) line is the contribution of the valence (conduction) band states. The factor 2 accounts two distinguished K-points of graphene.